\documentclass[aps,twocolumn,floatfix, showkeys, superscriptaddress, nofootinbib]{revtex4}
\usepackage{amsmath}
\usepackage{graphicx,bm,hhline}
\DeclareMathAlphabet{\pazocal}{OMS}{zplm}{m}{n}

\newcommand{\br}{{\bm r}}

\newcommand{\brp}{{\br}^\prime}

\newcommand{\hf}{\frac{1}{2}}

\newcommand{\wig}[1]{\mathrel{\hbox{\hbox to 0pt{\lower.6ex\hbox{$\sim$}\hss}\raise.4ex\hbox{$#1$}}}}

\makeatletter
\renewcommand*\env@matrix[1][\arraystretch]{%
  \edef\arraystretch{#1}%
  \hskip -\arraycolsep
  \let\@ifnextchar\new@ifnextchar
  \array{*\c@MaxMatrixCols c}}
\makeatother

\begin{document}
\title{Charge State Distributions in Dense Plasmas}

\author{J. White}
\affiliation{Los Alamos National Laboratory, P.O. Box 1663, Los Alamos, NM 87545, U.S.A.}

\author{W. Johns}
\affiliation{Los Alamos National Laboratory, P.O. Box 1663, Los Alamos, NM 87545, U.S.A.}

\author{C. J. Fontes}
\affiliation{Los Alamos National Laboratory, P.O. Box 1663, Los Alamos, NM 87545, U.S.A.}

\author{N. M. Gill}
\affiliation{Los Alamos National Laboratory, P.O. Box 1663, Los Alamos, NM 87545, U.S.A.}

\author{N. R. Shaffer}
\affiliation{Los Alamos National Laboratory, P.O. Box 1663, Los Alamos, NM 87545, U.S.A.}

\author{C. E. Starrett}
\email{starrett@lanl.gov}
\affiliation{Los Alamos National Laboratory, P.O. Box 1663, Los Alamos, NM 87545, U.S.A.}

\date{\today}
\begin{abstract}
Charge state distributions in hot, dense plasmas are a key ingredient in the calculation of spectral quantities like the opacity.  However, they are challenging to calculate, as models like Saha-Boltzmann become unreliable for dense, quantum plasmas.  Here we present a new variational model for the charge state distribution, along with a simple model for the energy of the configurations that includes the orbital relaxation effect.  Comparison with other methods reveals generally good agreement with average atom based calculations, the breakdown of the Saha-Boltzmann method, and mixed agreement with a chemical model.  We conclude that the new model gives a relatively inexpensive, but reasonably high fidelity method of calculating the charge state distribution in hot dense plasmas, in local thermodynamic equilibrium.
\end{abstract}
\pacs{ }
\keywords{Dense plasmas, Charge State Distributions}
\maketitle

\section{Introduction}
Recent and ongoing experiments have measured the opacity and related spectral quantities of dense plasmas at local thermodynamic equilibrium (LTE) conditions
\cite{ciricosta12,hoarty13,bailey15, nagayama19, perry17}.  Such experiments have highlighted possible weaknesses in modeling capability, and spurred theoretical interest \cite{pain17, pain20, more20, iglesias14, colgan16, fontes15, mondet15,krief16}.  Hot dense plasmas exist not only in the laboratory measurements of spectra, but also in stars \cite {blouin20} and in inertial fusion plasmas \cite {gomez14, perry17}, and the opacity is an important quantity for understanding energy transport in these systems.

Finite-temperature density functional theory (DFT) based models of hot dense plasmas have proven to be very useful for calculating material properties such as conductivity and equation of state \cite{knudson18, mazevet05, hu14}.  They can also be used to calculate opacity \cite{johnson06, shaffer17, hollebon19, hu14}.  However, due to the Fermi-Dirac occupation of states and the use of Kohn-Sham orbitals,
such calculations miss the important physics that arises from considering
the plasma to be composed of individual ions characterized by configurations
with integer occupation numbers. This physical picture naturally includes
the concept of a charge state distribution, which describes the total number
of ions in each ionization stage, i.e. those ions that possess the same integer
number of bound electrons. Furthermore, this picture allows for the calculation
of many more lines (radiative transitions) that appear in the measured spectra.

Several models have been presented that combine DFT-like treatments of free electrons with an integer-occupation treatment of bound orbitals \cite{blenski07, massacrier94, faussurier18, piron13, son14} to give the energies of the configurations.  Using these energies, a population model is then used to determine which of the configurations exist in the plasma and with what probability.
Alternative to these models are approaches that use isolated-ion electronic structure calculations, and then introduce the plasma effects, such as bound state pressure ionization, via a secondary model \cite{hakel04}.  

Here we present a new variational method, which uses ions screened by free electrons, to predict the charge state distribution.  Because of the free electron screening, the model is relatively simple to implement and is computationally cheap, but still includes the effect of orbital relaxation.  The variational model uses a constrained free energy minimization to determine the charge state distribution.  The key constraint is the average ionization, here provided by the {\texttt Tartarus} average atom model \cite{starrett18}.  It is argued that the use of free electron screening is justified (as opposed to solving the Schr\"odinger equation for the continuum states) since in either case there are more important physical approximations being made.  We compare predictions of the charge state distribution to other models, finding good agreement with other average atom based models that do not use this free electron approximation.  Disagreements with the ChemEOS model \cite{hakel04,hakel06, kilcrease15} are discussed.

\section {Theory }
\subsection{Variational Model for Charge State Distribution}
The physical model is of ions and an electron gas in spheres of equal volume V, given by the average volume per ion, and an overall average ionization $\bar{Z}$.  We require each sphere to be charge neutral, which should be a reasonable approximation for LTE plasmas.  If an ion corresponding to configuration $x$ has $Q_x$ bound electrons, the remaining electrons are considered to be free, with a density $n_x^0$, such that $Z_x=Q_x + \bar{Z}_x$, where $Z_x$ is the nuclear charge of the ion, and $\bar{Z}_x = V n_x^0$ is the number of ionized electrons for the ion.
The free energy $F$ per ion of a plasma composed of ions with a neutralizing electron gas is 
\begin{equation}
F= \sum_x W_x \left( U_x +T \ln W_x \right) + F^{eg}
\label{f1}
\end{equation}
where the sum is over all electronic configurations of the ions in the plasma,
and we are defining a configuration to mean a set of integer occupation numbers for orbitals defined by their principal, orbital angular, magnetic, and spin quantum numbers.
 $W_x$ is the probability that configuration $x$ exists, $T$ is the temperature of the plasma, $F^{eg}$ is the free energy of the electron gas \footnote{For which we take the average value, ignoring the variation from ion to ion.}, and  $U_x$ is the configuration average internal energy of configuration $x$.   We have used Hartree atomic units in which $\hbar =m_e = k_B = 1$.    

The entropy term in equation (\ref{f1}), $W_x\ln W_x$, is the entropy of mixing due to configuration $x$, and due to the bound electrons.  The occupations of electrons in bound orbitals, for a given configuration $x$, are an input to the model.  The complete list of the configurations to be included in the sum over $x$ is not known a priori, but is determined by an exhaustive search which we will discuss in the next section.  

While the bound orbital occupancy is fixed by the choice of configuration, in this model the free states are occupied according to Fermi-Dirac statistics.  This breaks the consistency between free and bound electrons so important for thermodynamic consistency of equation of state (EOS) models \cite{blenski95}, but it is of less importance for photon spectra due to the (generally) large number of configurations \cite{piron13}.  In average atom based EOS models a large amount of effort has been devoted to accurate treatment of the free orbitals \cite{blenski95, wilson06, scaalp, starrett18}.  However, for spectral resolution, an important physical effect completely missing in these single center models for dense plasmas is multiple scattering, which destroys the atomic eigenstate character of loosely bound orbitals \cite{starrett20ms}.  In light of this, it seems like unnecessary effort to apply the same level of rigor and numerical accuracy to calculating the free orbitals when solving for the ion's electronic structure.  Even if one were to solve for the continuum orbitals in the same potential as the bound orbitals, consistency with bound orbitals  would still be broken and important physical effects would still be missing.  

The main physical effect that we wish to capture is the influence of plasma screening on the ion's eigenvectors and eigenvalues.  This can be included by approximating the free electrons as a {\it truly} free electron gas, i.e. the free electron density is homogeneous, given by
\begin{equation}
n_x^0 = \frac{\sqrt{2}(T)^{3/2}}{\pi^2} F_{\hf}(\mu+\bar{V}_x,T)
\end{equation}
where $F_{\hf}$ is the Fermi integral (definition in reference \cite{starrett18}), $\mu$ is the electron chemical potential, and $\bar{V}_x$ is a local variation in potential taking a value such that the correct $\bar{Z}_x$ of ion $x$ is recovered.

To ensure that the correct average ionization $\bar{Z}$ for the plasma is recovered, a constrained minimization of the free energy is used.  Let $\Omega$ be the constrained free energy
\begin{equation}
\Omega  = F - B
\left(
\sum_xW_x -1
\right)- C
\left(
\sum_xW_x \bar{Z}_x-\bar{Z}
\right)
\end{equation}
where $B$ and $C$ are Lagrange multipliers.  The first constraint ensures that the probabilities sum to 1, the second ensures that the average ionization of the plasma is $\bar{Z}$.  Minimizing with respect to $W_x$ 
and enforcing the first constraint gives
\begin{equation}
W_x = g_x \frac{\exp\left(
-\frac{1}{T}\left(
U_x-C \bar{Z}_x
\right)
\right)}
{\Xi}
\end{equation}
where
\begin{equation}
\Xi = \sum_x g_x \exp\left(
-\frac{1}{T}\left(
U_x-C \bar{Z}_x
\right)
\right)
\end{equation}
is the partition function, and the sum is now over configurations with different energies $U_x$, with a configuration now being defined by a set of integer occupation numbers together with their principal and orbital angular momentum quantum numbers, and $g_x$ is the degeneracy of such a configuration.  The second constraint is enforced by varying $C$ until 
\begin{equation}
\sum_x W_x \bar{Z}_x = \bar{Z}
\label{cfind}
\end{equation}
is satisfied, which is done numerically.

To close this model, $\bar{Z}$ must be provided.  As there is no unique definition of this quantity, one has some freedom.  We have found that using an average atom model with a `chemical' definition works well.  In this  definition the number of bound electrons is the number of electrons in orbitals that asymptotically decay, and so are localized around the ion, consistent with the ion concept above.  We use the {\texttt Tartarus} average atom model of reference \cite{starrett18}.  In the notation of that reference, the chemical definition of ionization is ``$\bar{Z}$''.  Note that using the alternative definition of average ionization given in reference \cite{starrett18} (``$Z^*$''), 
is inconsistent with the present model's definition of an ``ion'', and in practice using $Z^*$ in place of $\bar{Z}$ sometimes causes equation (\ref{cfind}) to have no solution.

\subsection{Energy of an ion}
The configuration average energy of an ion, $U_x$, corresponding to an integer occupation number configuration $x$, can be approximated in the local density approximation (LDA) by the expression 
\begin{equation}
U_x = U^{el} _x+ \Delta U^{xc}_x + \Delta U^k_x 
\end{equation}
where $U^{el}_x$ is the electrostatic energy, $\Delta U^k_x$ is the kinetic energy of the electrons without a free electron contribution and $\Delta U^{xc}_x$ is the exchange and correlation internal energy, again with the free electron contribution removed.  $U_x^{el} $ is given by
\begin{equation}
U^{el}_x = \hf \int_V d^3r \left[ V^{el}_x(r) - \frac{Z_x}{r}\right] n_x(r)
\end{equation}
where the electrostatic potential is 
\begin{equation}
V^{el}_x(r) = - \frac{Z_x}{r} + \int_{V} d^3r' \frac{n_x(r')}{|\br - \brp|}
\end{equation}
and the ion electron density is
\begin{equation}
n_x(r) = \sum_{i}q_x^i \frac{P_{i,x}(r)^2}{4\pi r^2} +n_x^0
\end{equation}
where $P_{i,x}(r) $ is an eigenfunction of the radial Schr\"odinger equation, and the sum is over all bound eigenstates of the configuration, with integer occupation $q_x^i$, with $0\leq q_x^i \leq 2(2l_i+1)$, and $l_i$ is the orbital angular momentum quantum number of the eigenstate.

The kinetic energy of the electrons in the configuration is given by
\begin{equation}
\Delta U_x^k = \sum_i q_x^i N_x^i \epsilon_x^i - \int_V d^3r V^{eff}_x (r) n_x(r)
\end{equation}
where $\epsilon_x^i$ is the eigenenergy of eigenstate $i$ in configuration $x$, and 
\begin{equation}
 N_x^i = \int_V P_{i,x}(r)^2 d^3r
 \end{equation}
This normalization-like integral will be equal to $1$ for deeply bound orbitals, but will be less than $1$ for weakly bound orbitals whose eigenfunctions have not decayed to zero by the ion sphere radius $R$.

The exchange and correlation energy is given by
\begin{equation}
\Delta U_x^{xc,LDA} =  \int_V d^3r\left\{ u_{xc}\left[ n_x(r) \right] -u_{xc}[n_x^0]\right\}
\end{equation}
and we have used the Perdew-Zunger functional \cite{perdew81} for our calculations.    This results in an effective interaction potential
\begin{equation}
V_x^{eff}(r) = V_x^{el}(r) +V^{xc}[n_x(r)]
\end{equation}
with $V^{xc} $ the corresponding exchange and correlation potential.  Note that the eigenfunctions are normalized over all space and the potential $V^{eff}_x(r)$ is formally assumed to be constant outside the spheres for this purpose.

The configuration energies are improved by replacing the LDA energy for the bound orbitals with the Hartree-Fock interaction energy \cite{cowan_book}.  This provides a correction that improves configuration energies \cite{cowan_book}.  In this approximation the energy correction $\Delta U_x^{HF}$ is given by
\begin{equation}
\Delta U_x^{HF} = U^{HF}_x - U_x^{LDA}
\end{equation}
where
\begin{equation}
U_x^{LDA} = \hf \int_V d^3r \int_Vd^3r'
\frac{n_x^b(r)n_x^b(r')}{|\br-\brp |} +\Delta U_x^{xc,LDA}
\end{equation}
with $n_x^b(r) = n_x(r) - n_x^0$, and
\begin{equation}
U^{HF}_x = \hf \sum_{i,j} q_x^i(q_x^j - \delta_{i,j})V_{i,j}
\end{equation}
$V_{i,j}$ is the shell-shell interaction energy \cite{cowan_book}.  We also include the (generally small) correlation correction given in reference \cite{cowan_book}.  It is worth pointing out that this Hartree-Fock correction is not self-consistent as it uses the LDA eigenfunctions.
\begin{figure}
\begin{center}
\begin{tabular}{c}
\includegraphics[scale=0.3]{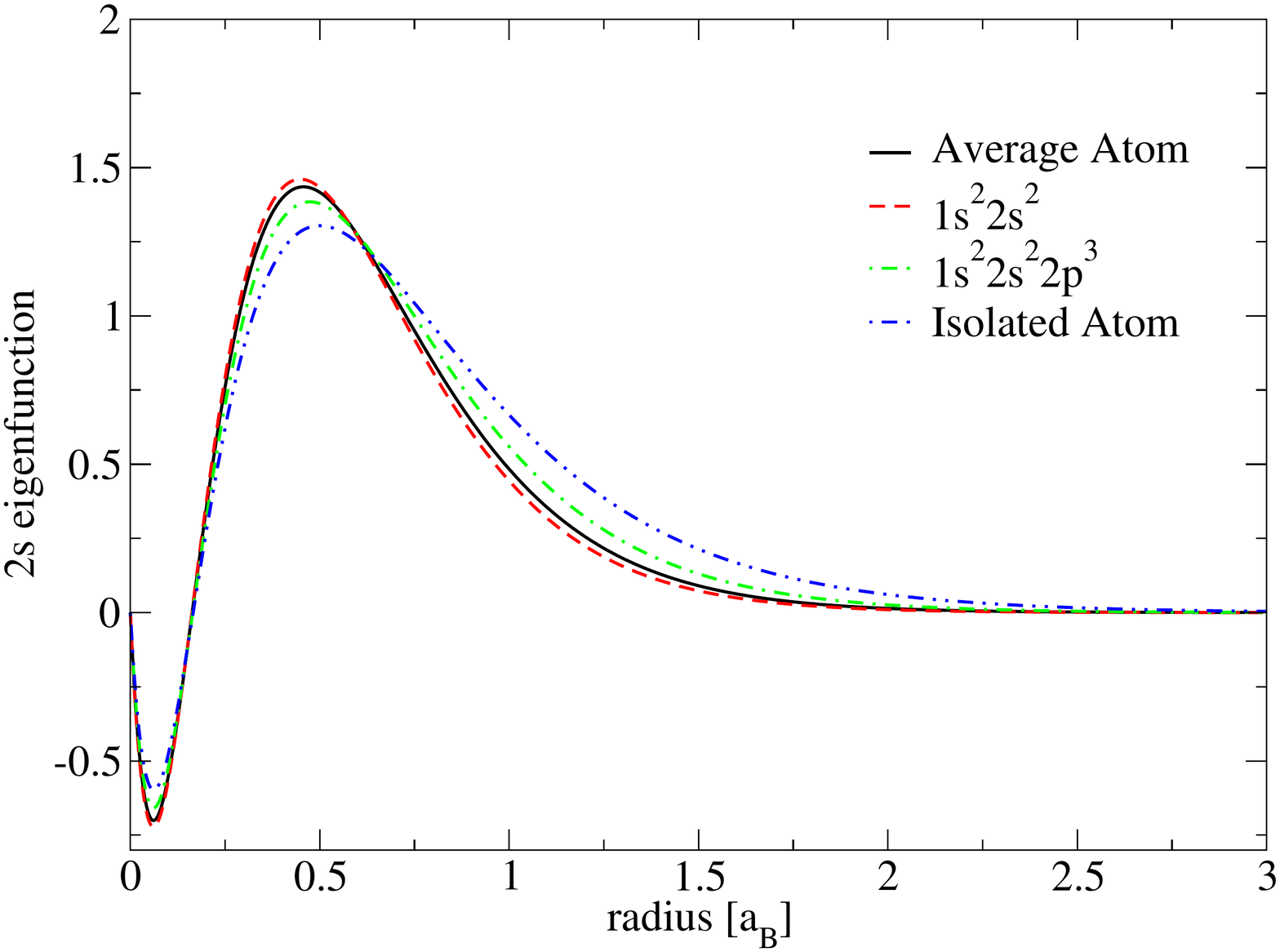} 
\end{tabular}
\end{center}
\caption{(Color online)  $2s$ eigenfunctions from the average atom model {\texttt Tartarus} ($\bar{Z}=7.9$) and from the present ion model with configurations $1s^22s^2$ ($\bar{Z}$=9) and $1s^22s^2 2p^3$ ($\bar{Z}=6$), at 100~eV and 2.7 g/cm$^3$.  Also shown is the $2s$ eigenfunction for a neutral, isolated aluminum atom.}
\label{fig_2s}
\end{figure}

\subsection {Discussion of some model approximations}
In comparison to other models of charge state distribution there are various differences.  A central difference from references \cite{faussurier18, piron13, blenski07}, which present models that also calculate density and temperature dependent eigenfunctions and eigenvalues, is that in those models the ion sphere sizes are varied from configuration to configuration so that the free electron density at the edge of the spheres is the same.  Here we allow those free electron densities to vary from ion to ion, while keeping the ion sphere volume and the chemical potential the same.  It does not seem probable that the amount of `space' an ion takes up will be strongly correlated with its ionization stage (at a given plasma density), as typical ionization/recombination times are much faster than ion motion time scales.  The concept of an ion sphere is itself a rather crude approximation in dense plasmas, with loosely bound valence states being strongly affected by neighboring ions \cite{starrett20ms}. Hence, such models' descriptions of weakly bound states can only be approximately correct in any case.  Therefore, our assumption of one ion sphere size is probably equally justified, while being simpler computationally, as there is no search for the unknown sphere sizes.

The ChemEOS model \cite{hakel04, hakel06, kilcrease15} starts from a different perspective.  It uses a calculation of eigenstates and eigenvalues of isolated ions.
The infinite bound spectrum of isolated ions leads to a divergent internal partition function and thus free energy. In ChemEOS, this is resolved with an ionization model that combines a hard-sphere free-volume model with a plasma microfield model for the destruction of bound orbitals.
In the present model the partition function converges because there is a finite list of configurations that exist.  This is due to the electronic structure being calculated self-consistently with the plasma effects, in contrast to the ChemEOS model.
\begin{figure}
\begin{center}
\begin{tabular}{c}
\includegraphics[scale=0.45]{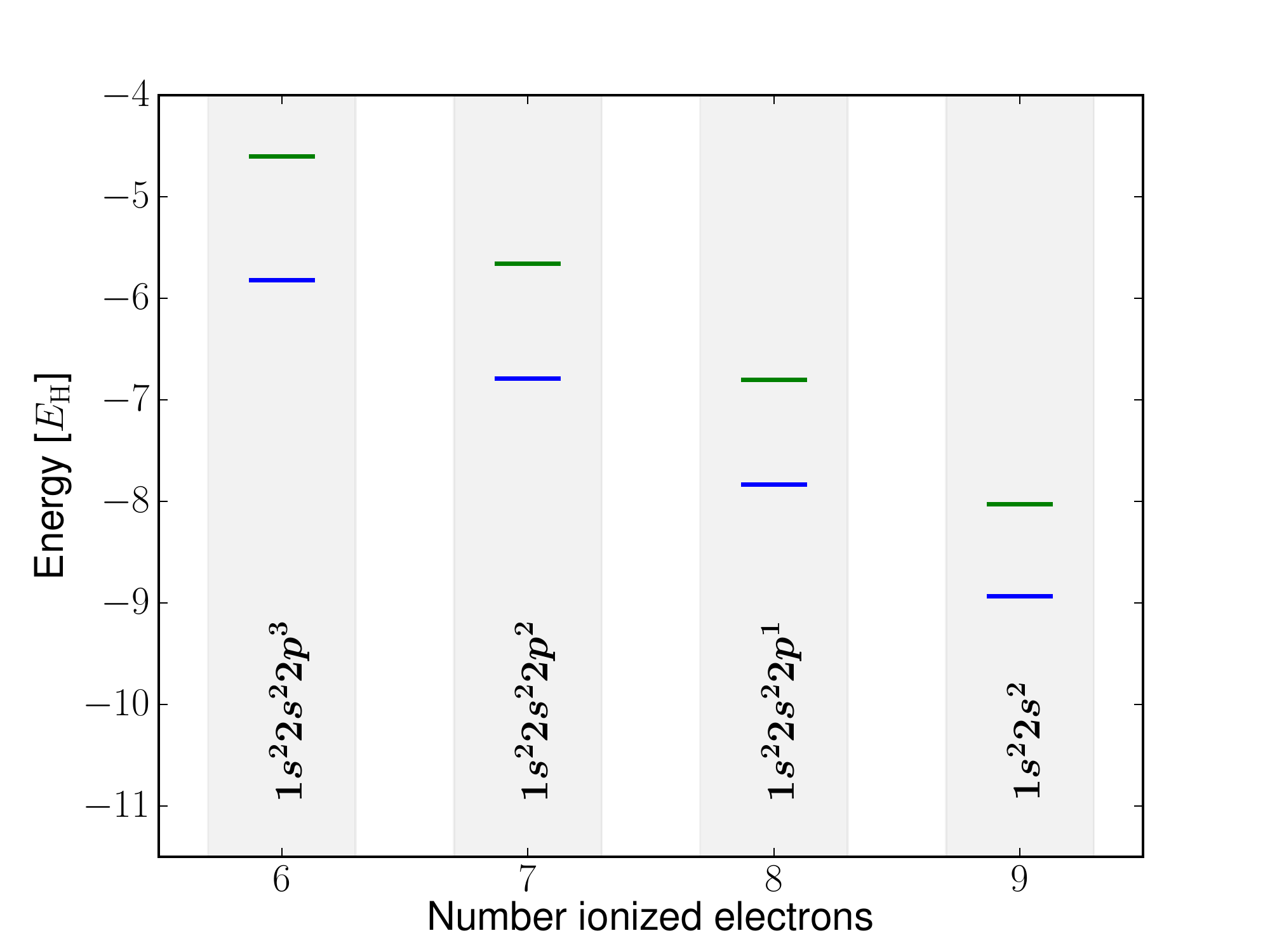} 
\end{tabular}
\end{center}
\caption{(Color online)  Eigenvalues for the $2s$ and $2p$ orbitals for the ground states of ion stages VII through X, for aluminum at solid density and 100 eV.  The upper green lines are for the $2p$ orbital, while the lower blue lines correspond to the more tightly bound $2s$ orbitals.  The fact that the $2s$ or $2p$ orbitals have different energies in different configurations is called orbital relaxation. }
\label{fig_or}
\end{figure}

In some models average atom eigenfunctions are used as approximations to the ion eigenfunctions \cite{piron13, wilson95}.  In figure \ref{fig_2s} the $2s$ eigenfunction from the average atom model {\texttt Tartarus}, for aluminum at 100 eV and 2.7 g/cm$^3$, is compared to that from two configurations, $1s^22s^2$ and $1s^22s^2 2p^3$, at the same temperature and density.  The average atom eigenfunction is quite close to the configuration wave functions, supporting the use of these wave functions as approximations for the configuration eigenfunctions \cite{piron13, wilson95}.  A disadvantage of such an approximation is that, because the average atom has a finite number of bound orbitals, if an excited state configuration is needed which requires an eigenfunction that does not exist in the average atom model, one does not have an approximate eigenfunction to use. 
By performing self-consistent field calculations for each configuration, the present model is able to recover excited-state orbitals which are pressure ionized out of the average atom, but which can appear in an integer-occupation ion due to orbital relaxation.

\begin{figure}
\begin{center}
\begin{tabular}{c}
\includegraphics[scale=0.45]{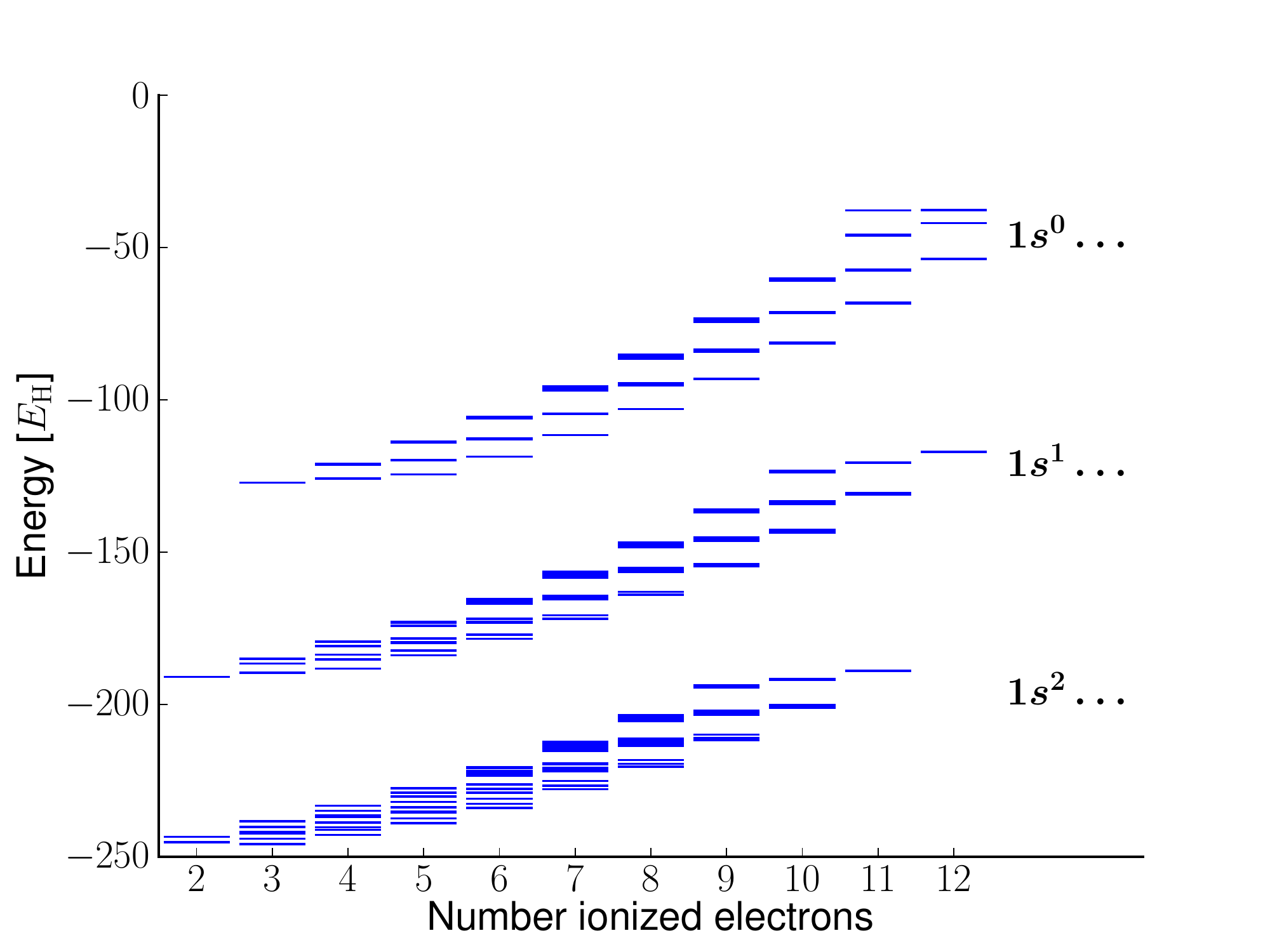} 
\end{tabular}
\end{center}
\caption{(Color online)  Energies of all singly or doubly excited ions  that exist in the present model, for ion stages III through XIII, for a solid density aluminum plasma at 100 eV.  The three distinct bands correspond to ions with the no excitations of the $1s$ orbital ($1s^2\ldots$), a single excitation ($1s^1\ldots$), or empty $1s$ orbital ($1s^0\ldots$).\label{fig_ie}}
\end{figure}

Also shown in figure \ref{fig_2s} is the $2s$ eigenfunction from a calculation of a neutral, ground state, isolated aluminum atom.  This is different from the other eigenfunctions due to density and temperature effects. While the eigenstate is well confined to within the ion sphere at 2.7 g/cm$^3$ (where the radius of the sphere is $\sim 3$ $a_B$), the screening due to other electrons changes from that due only to other bound electrons, to screening from free electrons as well as bound electrons.

The present configuration model includes the effect known as orbital relaxation, where the eigenvalue of a particular $nl$ orbital varies from configuration to configuration.  This is demonstrated in figure \ref{fig_or}, where the eigenvalues for the $2s$ and $2p$ orbitals are shown for the ground state configuration, for a range of ionization stages of aluminum.  Going to higher ionization, the orbitals become more tightly bound.  This is due to free electrons being less effective at screening the nucleus than bound electrons, as bound electrons are, on average, closer to the nucleus.  Note that the eigenvalue is shown for the $2p$ orbital of the $1s^22s^2$ configuration despite the fact that it is not occupied, and it does not directly affect the calculation of the energy of the ion.

In figure \ref{fig_ie} we show the energies of all configurations that exist for ion stages III through XIII, according to the present model, for a solid density aluminum plasma at 100 eV.  To determine whether or not an ion configuration exists, the model is run with the desired set of integer occupation numbers, and if a self-consistent field solution can be found for that configuration, then the configuration exists.  As one does not know the final list of configurations that exist a priori, many more configurations than those that are found to exist must be tested.  In the figure, three distinct bands are predicted, corresponding to a fully occupied, singly occupied, or empty $1s$ orbital.  In the {\texttt Tartarus} model the eigenenergy of the $1s$ orbital is $-61.6$ E$_H$, roughly equal to the gap between the observed bands.  Note that a fully ionized atom would have zero energy on this scale.  In the figure, ion stages with more ionized electrons have energies closer to this limit.
\begin{figure}
\begin{center}
\begin{tabular}{c}
\includegraphics[scale=0.3]{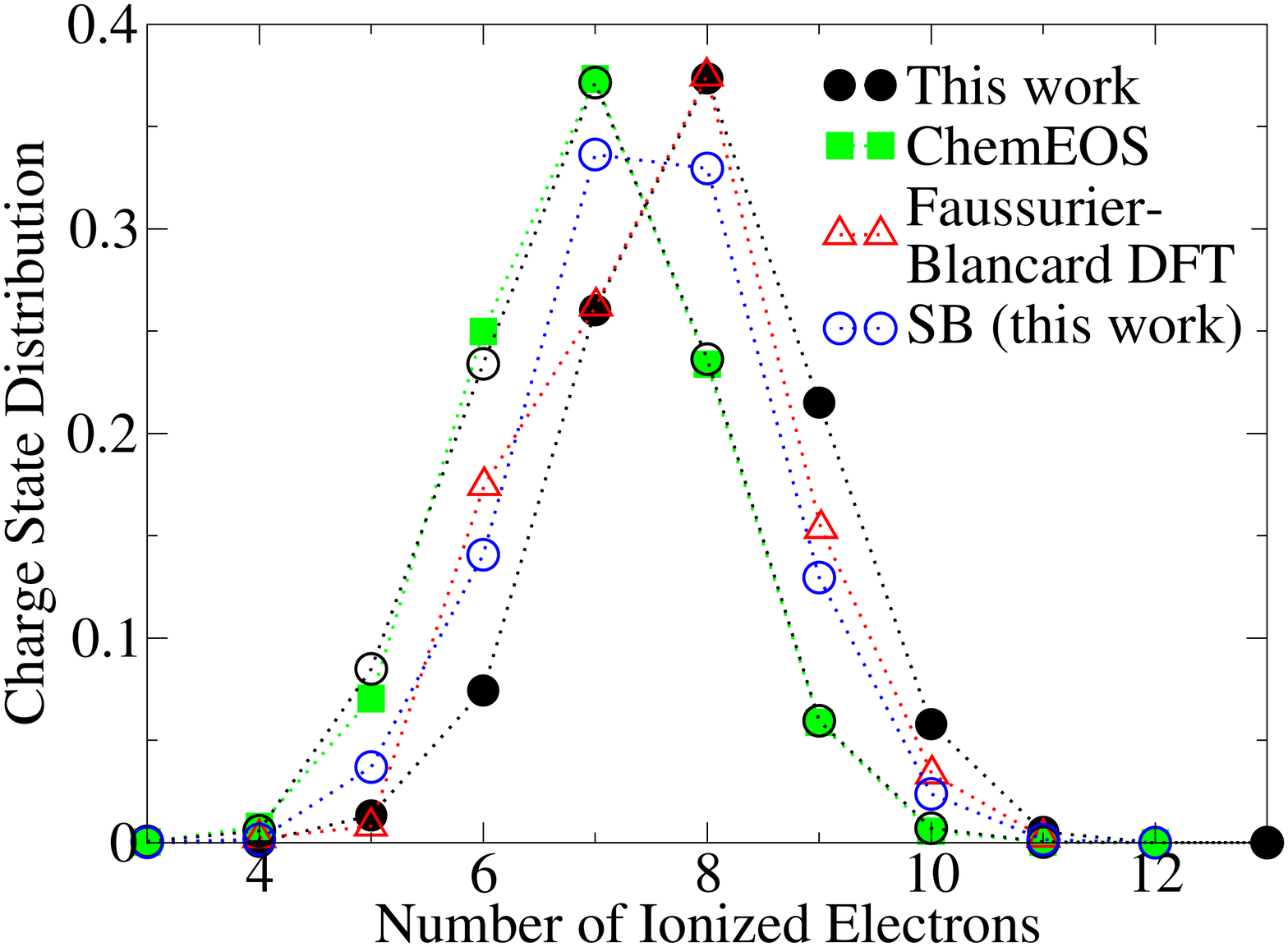} 
\end{tabular}
\end{center}
\caption{(Color online) Charge state distributions for aluminum at 100 eV and solid density.  Filled black circles are current variational model with HF energy for bound states, filled green squares are the ChemEOS model \cite{hakel04,hakel06,kilcrease15}, and red triangles are from reference \cite{faussurier18}, where we have used their model labeled `DFT'.  Also shown are open black circles which come from the present variational model, but using the ChemEOS value for the average ionization, and a Saha-Boltzmann (SB) calculation using the present ion configurations and energies.}
\label{fig_fauss}
\end{figure}

\section{results}
\begin{figure}
\begin{center}
\begin{tabular}{c}
\includegraphics[scale=0.4]{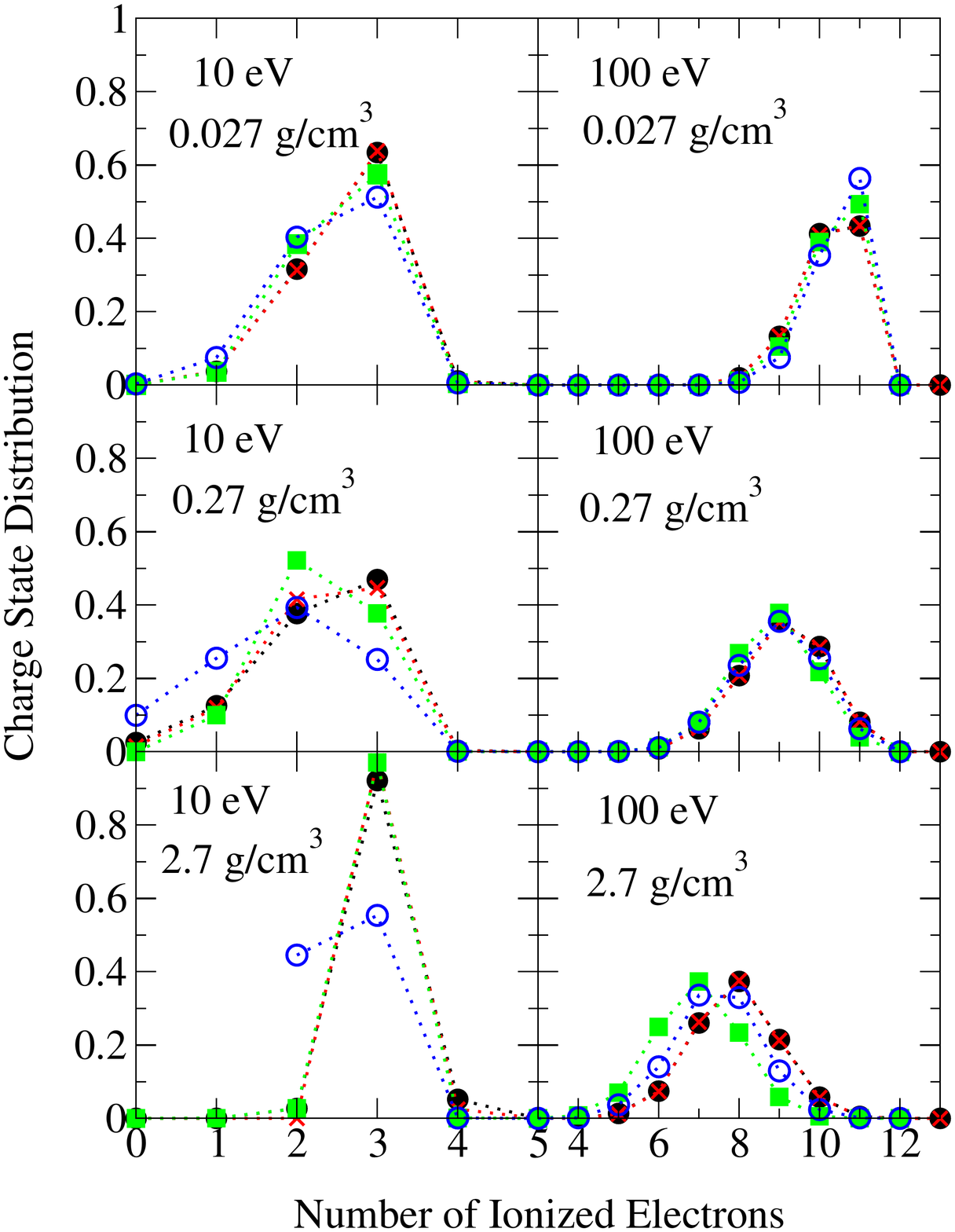} 
\end{tabular}
\end{center}
\caption{(Color online)  Charge state distributions for aluminum plasmas at various temperatures and densities.  Filled black circles are current variational model with HF energy for bound states, red crosses are current variational model with LDA  energy for bound states, open blue circles are from Saha-Boltzmann model with HF ionization energies from the present ion model, and filled green squares are the ChemEOS model.}
\label{fig_alcsd}
\end{figure}
In figure \ref{fig_fauss} we shown the charge state distribution for an aluminum plasma at solid density and a temperature of 100 eV.  Shown is the result from the present model compared to the recent model of Faussurier and Blancard \cite{faussurier18} (FB) and to the ChemEOS model \cite{hakel04, hakel06, kilcrease15}.  The FB model uses varying ion-sphere sizes to maintain the same free electron density at the edge of the spheres, has built in plasma screening, and uses a Saha-Boltzmann formalism to determine the charge state distribution and average ionization.  The agreement between the present model and the FB model is quite good.  The FB model corresponds to a lower average ionization ($\bar{Z}=7.60$) than the present model ($\bar{Z}=7.90$).  This may be due to the Saha-Boltzmann model, which is known to underestimate $\bar{Z}$ at high material densities because it is based on classical statistical mechanics.  We have also implemented the Saha-Boltzmann model using the ion configurations and energies of the present model.  This result is also shown in figure \ref{fig_fauss}, as open blue circles ($\bar{Z}=7.45$).  While it does not agree exactly with the FB model (due to the different ion models used), it is shifted to lower ionizations than the full variational model, indicating that the Saha-Boltzmann model is, at least partially, the cause of the lower average ionization.

Also shown in figure \ref{fig_fauss} is the result from the ChemEOS model.  It gives a significantly lower average ionization ($\bar{Z}=6.95$) than the present variational model ($\bar{Z}=7.90$), but the shape of the distribution is similar.   The present average ionization comes from the {\texttt Tartarus} average atom model, whereas $\bar{Z}$ for ChemEOS comes from the above mentioned hard-sphere and microfield model.  We have also used the ChemEOS $\bar{Z}$ to constrain the present variational model, and the result is also shown in figure \ref{fig_fauss}, as open black circles.  There is very close agreement with ChemEOS, indicating that the different methods of calculating the average ionization are indeed the cause of the differences, while the shape of the charge distributions is similar in both models for this case.

In figure \ref{fig_alcsd} charge state distributions for aluminum plasmas at temperatures of 10 and 100 eV, and three orders of magnitude in density, ranging from solid density to 1/100$^{th}$ of solid, are shown.  Shown are the present variational model using ion energies with and without the Hartree-Fock correction.  This correction makes little difference to any of the cases shown because it amounts to a roughly constant shift in configuration energy.  It should be important, however, when calculating spectra \cite{wilson95}.  Also shown in the figure are the results from the ChemEOS model and our own Saha-Boltzmann calculation. Agreement between the variational model and ChemEOS is generally very good, for all cases.  The exception is the 100 eV, solid density case discussed previously with figure \ref{fig_fauss}.  The Saha-Boltzmann model also compares well to the variational model at 100 eV, but is less reliable at the lower temperatures, and higher densities, as expected.

\begin{figure}
\begin{center}
\begin{tabular}{c}
\includegraphics[scale=0.4]{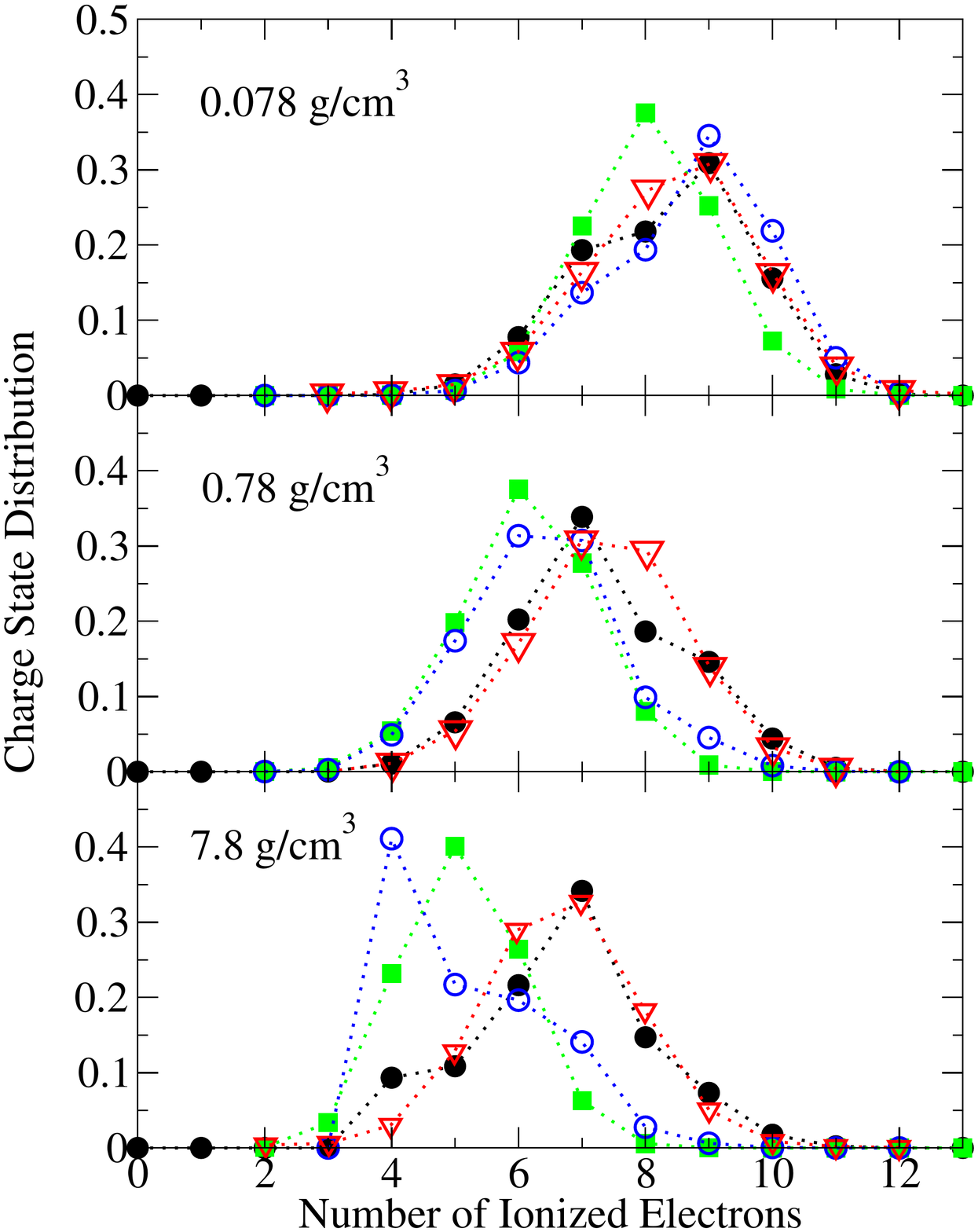} 
\end{tabular}
\end{center}
\caption{(Color online) Charge state distributions for iron plasmas at various densities a temperature of 40 eV.  Filled black circles are current variational model with HF energy for bound states, open blue circles are from Saha-Boltzmann model with HF ionization energies, filled green squares are the ChemEOS model, and open red triangles are from reference \cite{piron13}, where we have used the model labeled `eq.(43)'.}
\label{fig_fecsd}
\end{figure}

In figure \ref{fig_fecsd} we show charge state distributions for iron plasmas at a temperature of 40 eV.  We compare the present model to the model of reference \cite{piron13}, which is based on an average atom model with a variational formula for the charge state distributions.  That model uses the average atom eigenfunctions and the free state density for all ion configurations, with a varying ion-sphere size.  The level of agreement is quite good, and only small differences are observed.  Such differences would translate to small but visible effects in spectra.

Also shown in figure \ref{fig_fecsd} are the results of the ChemEOS model, which consistently predicts a lower average ionization than the present model, especially as density increases.  This is due to the very different way that $\bar{Z}$ is calculated.    In the present model  plasma effects are included self-consistently in the calculation of $\bar{Z}$, in contrast to the ChemEOS model.  Thus, we may expect the present model to be more realistic for this quantity where plasma effects are more significant.  However, as $\bar{Z}$ is not uniquely definable, we must be cautious before drawing strong conclusions.

Lastly, shown in figure \ref{fig_fecsd} is the result from our own calculation using the Saha-Boltzmann formalism.  At low density agreement between it and the full variational model is good, but gets  worse as density increases.  This is to be expected, as the Saha-Boltzmann model becomes steadily less reliable as density increases, for fixed temperature.

\section {Conclusions}
We have presented a new variational  model for calculating the charge state distribution in LTE plasmas.  The model is based on a free energy minimization, constrained to give the desired free electron density.  The inputs to the model are the energies of the configurations that exist, and the average ionization $\bar{Z}$.  For $\bar{Z}$ we use the average atom model {\texttt Tartarus}.  For the configuration energies we developed a free-electron screened ion model, which is used to determine whether a configuration exists in the plasma, and in which each configuration has a finite list of bound orbitals (occupied and unoccupied), due to free electron screening.  

Comparison with another variational model, reference \cite{piron13}, reveals good agreement.  An important difference with that model is that the present model includes orbital relaxation, which will be important for spectra calculations.  Comparison with the recent model of reference \cite{faussurier18} also results in good agreement.  That model uses the Saha-Boltzmann formalism, and using our own implementation of this, we demonstrated its breakdown at high densities and lower temperatures.  
Finally, comparison with the ChemEOS model \cite{hakel04,hakel06,kilcrease15} reveals mixed agreement.  Typically, agreement was better for lower densities, as expected.

\section*{Acknowledgments}
This work was performed under the auspices of the United States Department of Energy under contract DE-AC52-06NA25396.

\bibliographystyle{unsrt}
\bibliography{phys_bib}

\begin{thebibliography}{10}

\bibitem{ciricosta12}
O.~Ciricosta, S.~M. Vinko, H.-K. Chung, B.-I. Cho, C.~R.~D. Brown, T.~Burian,
  J.~Chalupsk\'y, K.~Engelhorn, R.~W. Falcone, C.~Graves, V.~H\'ajkov\'a,
  A.~Higginbotham, L.~Juha, J.~Krzywinski, H.~J. Lee, M.~Messerschmidt, C.~D.
  Murphy, Y.~Ping, D.~S. Rackstraw, A.~Scherz, W.~Schlotter, S.~Toleikis, J.~J.
  Turner, L.~Vysin, T.~Wang, B.~Wu, U.~Zastrau, D.~Zhu, R.~W. Lee, P.~Heimann,
  B.~Nagler, and J.~S. Wark.
\newblock Direct measurements of the ionization potential depression in a dense
  plasma.
\newblock {\em Phys. Rev. Lett.}, 109:065002, Aug 2012.

\bibitem{hoarty13}
D.~J. Hoarty, P.~Allan, S.~F. James, C.~R.~D. Brown, L.~M.~R. Hobbs, M.~P.
  Hill, J.~W.~O. Harris, J.~Morton, M.~G. Brookes, R.~Shepherd, J.~Dunn,
  H.~Chen, E.~Von~Marley, P.~Beiersdorfer, H.~K. Chung, R.~W. Lee, G.~Brown,
  and J.~Emig.
\newblock Observations of the effect of ionization-potential depression in hot
  dense plasma.
\newblock {\em Phys. Rev. Lett.}, 110:265003, Jun 2013.

\bibitem{bailey15}
James~E. Bailey, Taisuke Nagayama, Guillaume~Pascal Loisel, Gregory~Alan
  Rochau, C.~Blancard, J.~Colgan, Ph. Cosse, G.~Faussurier, C.~J. Fontes,
  F.~Gilleron, et~al.
\newblock A higher-than-predicted measurement of iron opacity at solar interior
  temperatures.
\newblock {\em Nature}, 517(7532):56--59, 2015.

\bibitem{nagayama19}
T.~Nagayama, J.~E. Bailey, G.~P. Loisel, G.~S. Dunham, G.~A. Rochau,
  C.~Blancard, J.~Colgan, Ph. Coss\'e, G.~Faussurier, C.~J. Fontes,
  F.~Gilleron, S.~B. Hansen, C.~A. Iglesias, I.~E. Golovkin, D.~P. Kilcrease,
  J.~J. MacFarlane, R.~C. Mancini, R.~M. More, C.~Orban, J.-C. Pain, M.~E.
  Sherrill, and B.~G. Wilson.
\newblock Systematic study of {L}-shell opacity at stellar interior
  temperatures.
\newblock {\em Phys. Rev. Lett.}, 122:235001, Jun 2019.

\bibitem{perry17}
T.~S. Perry, R.~F. Heeter, Y.~P. Opachich, P.~W. Ross, J.~L. Kline, K.~A.
  Flippo, M.~E. Sherrill, E.~S. Dodd, B.~G. DeVolder, T.~Cardenas, et~al.
\newblock Replicating the {Z} iron opacity experiments on the {NIF}.
\newblock {\em High Energy Density Physics}, 23:223--227, 2017.

\bibitem{pain17}
J.-C. Pain, F.~Gilleron, Q.~Porcherot, T.~Blenski, and Djamel Benredjem.
\newblock The hybrid detailed / statistical opacity code sco-rcg: New
  developments and applications.
\newblock {\em AIP Conference Proceedings}, 1811(1):190010, 2017.

\bibitem{pain20}
Jean-Christophe Pain and Franck Gilleron.
\newblock A quantitative study of some sources of uncertainty in opacity
  measurements.
\newblock {\em High Energy Density Physics}, 34:100745, 2020.

\bibitem{more20}
R.~More, J-C. Pain, S.~B. Hansen, T.~Nagayama, and J.~E. Bailey.
\newblock Free-free matrix-elements for two-photon opacity.
\newblock {\em High Energy Density Physics}, 34:100717, 2020.

\bibitem{iglesias14}
Carlos~A Iglesias.
\newblock A plea for a reexamination of ionization potential depression
  measurements.
\newblock {\em High Energy Density Physics}, 12:5--11, 2014.

\bibitem{colgan16}
James Colgan, David~P. Kilcrease, N.~H. Magee, Manolo~E. Sherrill,
  J.~Abdallah~Jr, Peter Hakel, Christopher~J. Fontes, Joyce~A. Guzik, and K.~A.
  Mussack.
\newblock A new generation of {L}os {A}lamos opacity tables.
\newblock {\em The Astrophysical Journal}, 817(2):116, 2016.

\bibitem{fontes15}
C.~J. Fontes, H.~L. Zhang, J.~Abdallah~Jr, R.~E.~H. Clark, D.~P. Kilcrease,
  J.~Colgan, R.~T. Cunningham, P.~Hakel, N.~H. Magee, and M.~E. Sherrill.
\newblock The {L}os {A}lamos suite of relativistic atomic physics codes.
\newblock {\em Journal of Physics B: Atomic, Molecular and Optical Physics},
  48(14):144014, 2015.

\bibitem{mondet15}
Guillaume Mondet, Christophe Blancard, Philippe Coss{\'e}, and G{\'e}rald
  Faussurier.
\newblock Opacity calculations for solar mixtures.
\newblock {\em The Astrophysical Journal Supplement Series}, 220(1):2, 2015.

\bibitem{krief16}
M.~Krief, A.~Feigel, and D.~Gazit.
\newblock Line broadening and the solar opacity problem.
\newblock {\em The Astrophysical Journal}, 824(2):98, 2016.

\bibitem{blouin20}
Simon Blouin, Nathaniel~R. Shaffer, Didier Saumon, and Charles~E. Starrett.
\newblock New conductive opacities for white dwarf envelopes.
\newblock {\em The Astrophysical Journal}, 899(1):46, aug 2020.

\bibitem{gomez14}
M.~R. Gomez, S.~A. Slutz, A.~B. Sefkow, D.~B. Sinars, K.~D. Hahn, S.~B. Hansen,
  E.~C. Harding, P.~F. Knapp, P.~F. Schmit, C.~A. Jennings, T.~J. Awe,
  M.~Geissel, D.~C. Rovang, G.~A. Chandler, G.~W. Cooper, M.~E. Cuneo, A.~J.
  Harvey-Thompson, M.~C. Herrmann, M.~H. Hess, O.~Johns, D.~C. Lamppa, M.~R.
  Martin, R.~D. McBride, K.~J. Peterson, J.~L. Porter, G.~K. Robertson, G.~A.
  Rochau, C.~L. Ruiz, M.~E. Savage, I.~C. Smith, W.~A. Stygar, and R.~A. Vesey.
\newblock Experimental demonstration of fusion-relevant conditions in
  magnetized liner inertial fusion.
\newblock {\em Phys. Rev. Lett.}, 113:155003, Oct 2014.

\bibitem{knudson18}
M.~D. Knudson, M.~P. Desjarlais, M.~Preising, and R.~Redmer.
\newblock Evaluation of exchange-correlation functionals with multiple-shock
  conductivity measurements in hydrogen and deuterium at the
  molecular-to-atomic transition.
\newblock {\em Phys. Rev. B}, 98:174110, Nov 2018.

\bibitem{mazevet05}
S.~Mazevet, M.~P. Desjarlais, L.~A. Collins, J.~D. Kress, and N.~H. Magee.
\newblock Simulations of the optical properties of warm dense aluminum.
\newblock {\em Physical Review E}, 71(1):016409, 2005.

\bibitem{hu14}
S.~X. Hu, L.~A. Collins, T.~R. Boehly, J.~D. Kress, V.~N. Goncharov, and
  S.~Skupsky.
\newblock First-principles thermal conductivity of warm-dense deuterium plasmas
  for inertial confinement fusion applications.
\newblock {\em Phys. Rev. E}, 89:043105, Apr 2014.

\bibitem{johnson06}
W.~R. Johnson, C.~Guet, and G.~F. Bertsch.
\newblock Optical properties of plasmas based on an average-atom model.
\newblock {\em Journal of Quantitative Spectroscopy and Radiative Transfer},
  99(1-3):327--340, 2006.

\bibitem{shaffer17}
N.R. Shaffer, N.G. Ferris, J.~Colgan, D.P. Kilcrease, and C.E. Starrett.
\newblock Free-free opacity in dense plasmas with an average atom model.
\newblock {\em High Energy Density Physics}, 23:31 -- 37, 2017.

\bibitem{hollebon19}
P.~Hollebon, O.~Ciricosta, M.~P. Desjarlais, C.~Cacho, C.~Spindloe,
  E.~Springate, I.~C.~E. Turcu, J.~S. Wark, and S.~M. Vinko.
\newblock Ab initio simulations and measurements of the free-free opacity in
  aluminum.
\newblock {\em Phys. Rev. E}, 100:043207, Oct 2019.

\bibitem{blenski07}
T.~Blenski and B.~Cichocki.
\newblock Variational approach to the average-atom-in-jellium and
  superconfigurations-in-jellium models with all electrons treated
  quantum-mechanically.
\newblock {\em High Energy Density Physics}, 3(1–2):34 -- 47, 2007.
\newblock Radiative Properties of Hot Dense Matter.

\bibitem{massacrier94}
Gérard Massacrier.
\newblock Self-consistent schemes for the calculation of ionic structures and
  populations in dense plasmas.
\newblock {\em Journal of Quantitative Spectroscopy and Radiative Transfer},
  51(1):221 -- 228, 1994.
\newblock Special Issue Radiative Properties of Hot Dense Matter.

\bibitem{faussurier18}
G{\'e}rald Faussurier and Christophe Blancard.
\newblock Density effects on electronic configurations in dense plasmas.
\newblock {\em Physical Review E}, 97(2):023206, 2018.

\bibitem{piron13}
R.~Piron and T.~Blenski.
\newblock Variational average-atom in quantum plasmas (vaaqp)--application to
  radiative properties.
\newblock {\em High Energy Density Physics}, 9(4):702--710, 2013.

\bibitem{son14}
Sang-Kil Son, Robert Thiele, Zoltan Jurek, Beata Ziaja, and Robin Santra.
\newblock Quantum-mechanical calculation of ionization-potential lowering in
  dense plasmas.
\newblock {\em Phys. Rev. X}, 4:031004, Jul 2014.

\bibitem{hakel04}
Peter Hakel and David~P Kilcrease.
\newblock {CHEMEOS}: A new chemical-picture-based model for plasma
  equation-of-state calculations.
\newblock In {\em AIP Conference Proceedings}, volume 730, pages 190--199.
  American Institute of Physics, 2004.

\bibitem{starrett18}
C.~E. Starrett.
\newblock High-temperature electronic structure with the
  {K}orringa-{K}ohn-{R}ostoker green's function method.
\newblock {\em Phys. Rev. E}, 97:053205, May 2018.

\bibitem{hakel06}
P.~Hakel, M.~E. Sherrill, S.~Mazevet, J.~Abdallah~Jr., J.~Colgan, D.~P.
  Kilcrease, N.~H. Magee, C.~J. Fontes, and H.~L. Zhang.
\newblock The new los alamos opacity code {ATOMIC}.
\newblock {\em Journal of Quantitative Spectroscopy and Radiative Transfer},
  99(1-3):265--271, 2006.

\bibitem{kilcrease15}
D.~P. Kilcrease, J.~Colgan, P.~Hakel, C.~J. Fontes, and M.~E. Sherrill.
\newblock An equation of state for partially ionized plasmas: The {C}oulomb
  contribution to the free energy.
\newblock {\em High Energy Density Physics}, 16:36--40, 2015.

\bibitem{blenski95}
Thomas Blenski and Kenichi Ishikawa.
\newblock Pressure ionization in the spherical ion-cell model of dense plasmas
  and a pressure formula in the relativistic {P}auli approximation.
\newblock {\em Physical Review E}, 51(5):4869, 1995.

\bibitem{wilson06}
B.~Wilson, V.~Sonnad, P.~Sterne, and W.~Isaacs.
\newblock Purgatorio--a new implementation of the inferno algorithm.
\newblock {\em J. Quant. Spect. Rad. Trans.}, 99:658, 2006.

\bibitem{scaalp}
G\'erald Faussurier, Christophe Blancard, Philippe Coss\'e, and Patrick
  Renaudin.
\newblock Equation of state, transport coefficients, and stopping power of
  dense plasmas from the average-atom model self-consistent approach for
  astrophysical and laboratory plasmas.
\newblock {\em Physics of Plasmas}, 17(5), 2010.

\bibitem{starrett20ms}
Charles~E. Starrett and Nathaniel Shaffer.
\newblock Multiple scattering theory for dense plasmas.
\newblock {\em arXiv preprint arXiv:2008.11616}, 2020.

\bibitem{perdew81}
J.~P. Perdew and Alex Zunger.
\newblock Self-interaction correction to density-functional approximations for
  many-electron systems.
\newblock {\em Phys. Rev. B}, 23:5048--5079, May 1981.

\bibitem{cowan_book}
Robert~D. Cowan.
\newblock {\em The {T}heory of {A}tomic {S}tructure and {S}pectra}.
\newblock Univ of California Press, 1981.

\bibitem{wilson95}
Brian~G. Wilson, David~A. Liberman, and Paul~T. Springer.
\newblock A deficiency of local density functionals for the calculation of
  self-consistent field atomic data in plasmas.
\newblock {\em Journal of Quantitative Spectroscopy and Radiative Transfer},
  54(5):857--878, 1995.

\end{thebibliography}

\end{document}